\documentclass[12pt]{article}
\begin{document}
\begin{center}

{\bf VISCOUS FRW COSMOLOGY IN MODIFIED GRAVITY}

\vspace{1cm}

I. Brevik\footnote{Email: iver.h.brevik@ntnu.no; corresponding
author.}, O. Gorbunova\footnote{On leave from Tomsk State
Pedagogical University, Tomsk, Russia. Email:
gorbunovaog@yandex.ru}, and Y. A. Shaido\footnote{On leave from
Tomsk State Pedagogical University, Tomsk, Russia. Email:
shaido@ngs.ru}

\bigskip

Department of Energy and Process Engineering, Norwegian University
of Science and Technology, N-7491 Trondheim, Norway

\bigskip

\today
\end{center}

\begin{abstract}
We discuss a modified form of gravity implying that the action
contains a power $\alpha$ of the scalar curvature. Coupling with
the cosmic fluid is assumed. As equation of state for the fluid,
we take the simplest version where the pressure is proportional to
the density. Based upon a natural ansatz for the time variation of
the scale factor, we show that the equations of motion are
satisfied for general $\alpha$. Also  the condition of
conservation of energy and momentum is satisfied. Moreover, we
investigate the case where the fluid is allowed to possess a bulk
viscosity, and find the noteworthy fact that consistency of the
formalism requires the bulk viscosity to be proportional to the
power $(2\alpha -1)$ of the scalar expansion. In Einstein's
gravity, where $\alpha=1$, this means that the bulk viscosity is
proportional to the scalar expansion. This mathematical result is
of physical interest; as discussed recently by the present
authors, there exists in principle a viscosity-driven transition
of the fluid from the quintessence region into the phantom region,
implying a future Big Rip singularity.
\end{abstract}

\section{Introduction}

Recent years have witnessed a considerable interest in modified
versions of Einstein's gravity. the motivation for this is coming
largely from the evidence that has accumulated about the ongoing
accelerated expansion of the universe. This evidence is seen in
the high redshifts of type Ia supernovae \cite{reiss98}, as well
as in the anisotropy power spectrum of the cosmic microwave
background \cite{bennet03}. These observations lead naturally to
the idea of a dark energy, believed to be  73\% of the total
energy content of the universe. Thermodynamically, the dark energy
concept implies rather bizarre consequences; thus the equation of
state parameter $w$ of the cosmic fluid may be less than -1,
implying  a singularity in the future called a Big Rip
 \cite{caldwell03}, and it leads also to negative
entropies \cite{brevik04}. In the papers listed in
Ref.~\cite{caldwell03}, actually two different types of Big Rip
were investigated. In the recent paper of Nojiri {\it et al.}
\cite{nojiri05}, a classification of the known types of Big Rip
were made and two new types of Big Rip were discovered, so that we
know now four types in all.

The intention of introducing modified gravity models is that one
may obtain a gravitational alternative to the conventional
description of dark energy. The possibility may look attractive as
the presence of dark energy may thereby be a consequence of the
expansion of the universe \cite{abdalla05}. Various aspects of the
modified gravity idea  have recently been  discussed in
Refs.~\cite{nojiri03,nojiri04,carroll04,shaido05} (for an
application to the de Sitter space, see Ref.~\cite{cognola05}.) In
the present paper we shall study the following modified gravity
model:
\begin{equation}
S=\frac{1}{2\kappa^2}\int d^4x\,\sqrt{-g}\left(f_0R^\alpha
+L_m\right). \label{1}
\end{equation}
Here $\kappa^2=8\pi G$, $f_0$ and $\alpha$ are constants ($\alpha$
may in principle be negative), and $L_m$ is the matter Lagrangian.
This kind of system was recently studied by Abdalla \makebox{{\it
et al.} \cite{abdalla05}}. Note that the dimension of $f_0$ is
$\rm{cm}^{2(\alpha-1)}$.

It is interesting to note that the string/M-theory effective
action may produce terms with negative powers of $R$
\cite{nojiri03a}. Now, modified gravity with an $1/R$ is in itself
not stable \cite{dolgov03}, but the inclusion of a $R^2$ term in
the action \cite{nojiri03}, or the inclusion of quantum effects
\cite{nojiri04a}, makes the theory stable. As we are interested in
the late time dynamics of our theory we may drop the $R^2$ terms
and work only with terms that are dominant at late times.

It turns out that even this apparently simple generalization of
Einstein's theory is not quite trivial. We will focus attention on
the following two points:

 $\bullet$ How the equations of motion as derived from the action
 (\ref{1}) are compatible with a scale factor $a$ varying with
 time according to the ansatz of Eq.~(\ref{6}) below. The fluid is
 in the first part of our work  taken to be non-viscous, and we adopt for simplicity the
 conventional equation of state
 \begin{equation}
 p=(\gamma -1)\rho= w\rho. \label{2}
 \end{equation}
 It turns out that the equations of motion are actually satisfied,
  for arbitrary values of $\alpha$.  Moreover, the
 conservation equation for  energy and momentum are  satisfied
 also. This is reassuring, as regards the adaptability of the
 conventional formalism to the modified gravity theory.

 $\bullet$ Our second point is to introduce a bulk viscosity $\zeta$ in
 the fluid. It turns out that in order to obtain a consistent
 description, one arrives in a natural way at a time dependent bulk viscosity
 model; see
 Eq.~(\ref{29}) below. In particular, in the case of Einstein's gravity ($\alpha=1$),
  the bulk viscosity becomes
  simply proportional to the scalar expansion. An interesting fact is
  that this is
precisely the relationship recently found in
 Ref.~\cite{brevik05} to be compatible with a viscosity-driven
 transfer of the fluid from the
  quintessence region ($w>-1$) into the phantom region ($w<-1$).

  \section{Solutions of the Equations of Motion. Non-Viscous Case}

  We start from the spatially flat FRW metric,
  \begin{equation}
  ds^2= -dt^2+a^2(t)\,d{\bf x}^2, \label{3}
  \end{equation}
  and set the cosmological constant $\Lambda$ equal to zero. The
  Hubble parameter is $H=\dot{a}/a$, where an overdot means
  differentiation with respect to $t$. The scalar expansion is
  $\theta={U^\mu}_{;\mu}=3H$, where $U^\mu$ is the four-velocity of
  the fluid.

We shall be primarily interested in the development of the late
universe, from $t=t_0$ onwards. For simplicity we choose the
origin of time such that $t_0=0$. The corresponding energy density
is $\rho_0$; this quantity is assumed to be known. Similarly we
assume the initial scalar expansion $\theta_0$ to be known. In
Einstein's gravity \cite{brevik05,brevik94}
\begin{equation}
\rho= \rho_0\left(1+\frac{1}{2}\gamma \theta_0 t\right)^{-2},
\label{4}
\end{equation}
\begin{equation}
a=a_0\left( 1+\frac{1}{2}\gamma \theta_0t\right)^{2/3\gamma}.
\label{5}
\end{equation}
In the recent modified gravity model of Abdalla {\it et al.}
\cite{abdalla05}, the exponent $2/3\gamma$ in Eq.~(\ref{5}) is
replaced by $2\alpha/3\gamma$. It becomes thus natural to assume
the following time variation of the scale factor:
\begin{equation}
a=a_0\left(1+\beta t\right)^{2\alpha/3\gamma} \label{6}
\end{equation}
 in the present case. We shall adopt this form in the following.
 The parameter $\beta$ has to be determined from the equations of
 motion. Taking the (00)-component of these equations, it is in
 principle possible to relate $\beta$ to the initial energy
 density $\rho_0$ at $t=0$. Instead of Eq.~(\ref{6}), we may
 equally well write
 \begin{equation}
 a=a_0\left( 1+\frac{\gamma}{2\alpha}\theta_0 \,t\right)^{2\alpha/3\gamma}, \label{7}
 \end{equation}
 $\theta_0=3H_0$ being the scalar expansion at $t=0$. The explicit
 time dependence of $\theta$ thus follows at once,
 \begin{equation}
 \theta=\theta_0\left( 1+\frac{\gamma}{2\alpha}\theta_0 \,t\right)^{-1}.
 \label{8}
 \end{equation}
 We see that $\beta$ and $\theta_0$ are equivalent parameters to
 be determined from the equations of motion as they are related
 through a constant factor,
 \begin{equation}
 \beta=\frac{\gamma}{2\alpha}\theta_0. \label{9}
 \end{equation}
 In Einstein's gravity,
 \begin{equation}
 \beta=\frac{1}{2}\gamma \theta_0, \quad \alpha =1. \label{10}
 \end{equation}
 In this case we find from Riemann's energy equation that $\theta$
 is related to $\rho$ in a very simple way,
 \begin{equation}
 \theta^2=3\kappa^2\rho; \label{11}
 \end{equation}
 cf., for instance, Refs.~\cite{brevik05,brevik94}. Equation
 (\ref{11}) holds for all values of $t$.

From the action (\ref{1}) we obtain in the general case the
equations of motion\cite{abdalla05}
\begin{eqnarray}
 -\frac{1}{2}f_0\,g_{\mu\nu}R^\alpha+\alpha
f_0\,R_{\mu\nu}R^{\alpha-1}-\alpha f_0\,\nabla_\mu\nabla_\nu
R^{\alpha-1} \nonumber \\
+\alpha f_0\,g_{\mu\nu}\nabla^2 R^{\alpha-1}=\kappa^2 T_{\mu\nu},
\label{12}
\end{eqnarray}
where $T_{\mu\nu}$ is the energy-momentum tensor corresponding to
$L_m$ (there is a printing error before Eq.~(3.2) in
Ref.~\cite{abdalla05}). The notation is chosen such that the
values $\alpha=1,\,f_0=1$ correspond to Einstein's gravity.

Let us put  $\mu=\nu=0$. In coordinate basis we have
\begin{equation}
R_{00}=-\frac{3\ddot{a}}{a},\quad
R=6\left(\frac{\ddot{a}}{a}+\frac{\dot{a}^2}{a^2}\right),\label{13}
\end{equation}

\begin{equation}
\nabla_0^2 R^{\alpha-1}=(\alpha-1)\partial_0(R^{\alpha-2}
\dot{R}),\label{14}
\end{equation}
\begin{equation}
\nabla^2 R^{\alpha -1}=-\frac{\alpha-1}{a^3}\partial_0(a^3
R^{\alpha-2}\dot{R}). \label{15}
\end{equation}
Then, since $T_{00}=\rho$, the (00)-component of the equations of
motion becomes
\begin{eqnarray}
\frac{1}{2}f_0 R^\alpha-\alpha
f_0\frac{3\ddot{a}}{a}R^{\alpha-1}-f_0\alpha(\alpha-1)\partial_0(R^{\alpha-2}\dot{R})
\nonumber \\
 -\alpha
f_0\frac{\alpha-1}{a^3}\partial_0(a^3R^{\alpha-2}\dot{R})=\kappa^2\rho.
\label{16}
\end{eqnarray}
Now taking into account our ansatz (\ref{6}) for the scale factor
we calculate
\begin{equation}
R=\frac{4\alpha \beta^2/\gamma}{(1+\beta
t)^2}\left(\frac{4\alpha}{3\gamma}-1\right),
 \label{17}
\end{equation}
\begin{equation}
\frac{3\ddot{a}}{a}R^{\alpha-1}=\frac{(2\beta)^{2\alpha}(2\alpha/3\gamma-1)}{2(1+\beta
t)^{2\alpha}}
\frac{\alpha}{\gamma}\left[\frac{\alpha}{\gamma}\left(\frac{4\alpha}{3\gamma}-1\right)\right]^{\alpha-1},
 \label{18}
\end{equation}
\begin{equation}
\nabla_0^2
R^{\alpha-1}=\frac{(2\beta)^{2\alpha}(\alpha-1)(2\alpha-1)}{2(1+\beta
t)^{2\alpha}}\left[\frac{\alpha}{\gamma}\left(\frac{4\alpha}{3\gamma}-1\right)\right]^{\alpha-1},
\label{19}
\end{equation}
\begin{equation}
\nabla^2
R^{\alpha-1}=\frac{-(2\beta)^{2\alpha}(\alpha-1)(2\alpha-2\alpha/\gamma-1)}{2(1+\beta
t)^{2\alpha}}\left[\frac{\alpha}{\gamma}\left(\frac{4\alpha}{3\gamma}-1\right)\right]^{\alpha-1},
\label{20}
\end{equation}
whereby we arrive at the equation
\begin{equation}
\frac{f_0\,\theta_0^{2\alpha}\,(\gamma/\alpha)^\alpha(4\alpha/3\gamma-1)^{\alpha-1}}
{2(1+\beta
t)^{2\alpha}}\Big\{(2-\alpha)\frac{2\alpha}{3\gamma}-(\alpha-1)(2\alpha-1)\Big\}
=\kappa^2 \rho, \label{21}
\end{equation}
reinstalling $\theta_0$ by using Eq.~(\ref{9}).  As is seen  from
Eq.~(\ref{21}), the energy density varies with time as
\begin{equation}
\rho=\frac{\rho_0}{(1+\beta t)^{2\alpha}}. \label{22}
\end{equation}
If the parameter set $\{f_0, \alpha, \rho_0\}$ is given, the value
of the initial scalar expansion $\theta_0$ can in principle be
determined from Eq.~(\ref{21}), setting $t=0$. Equation (\ref{21})
agrees with Ref.~\cite{abdalla05} (the notation, and the initial
assumption (\ref{6}) for the scale factor, are different).

The case of Einstein's gravity, $\alpha=1,\,f_0=1$, yields
\begin{equation}
\theta_0^2=3\kappa^2 \rho_0, \label{23}
\end{equation}
which is in accordance with Eq.~(\ref{11}) when $t=0$. It is seen
that $\theta^2$ and $\rho$ are  both varying with time as
$(1+\beta t)^{-2}$.

Next, we may for general $\alpha$ put $\mu=\nu=r$ in
Eq.~(\ref{12}), to calculate the pressure component of the
equations of motion. We have, in coordinate basis,
$T_{rr}=pg_{rr}=pa^2$. We abstain, however, from giving the
details here. The most important point in our context is that the
pressure is from this equation found to vary with time in the same
manner as the density $\rho$,
\begin{equation}
p=\frac{p_0}{(1+\beta t)^{2\alpha}}. \label{24}
\end{equation}
Thus, if the equation of state (\ref{2}) holds at $t=0$, it holds
for all later values of $t$ also, with the same constant values
for $\gamma$ and $w$. This is important for the consistency and
applicability of the  modified gravity theory.

\subsection{A remark on energy-momentum conservation}

So far, we have not taken into account the condition coming from
conservation of energy and momentum. A characteristic property
shows up if one takes the covariant divergence on both sides of
Eq.~(\ref{12}): terms on the left hand side of the equation
combine to yield {\it zero}. This is shown explicitly in a recent
calculation of Koivisto \cite{koivisto05}. Thus we obtain the same
conservation equation for energy-momentum as in ordinary
relativity:
\begin{equation}
\nabla^\nu\,T_{\mu\nu}=0, \label{25}
\end{equation}
which in turn implies
\begin{equation}
\frac{d\rho}{dt}=-\frac{\rho +p}{a^3}\frac{d}{dt}a^3. \label{26}
\end{equation}
Inserting the forms (\ref{22}) and (\ref{24}) for $\rho$ and $p$
we see that Eq.~(\ref{26}) is satisfied identically. Again, this
is a point supporting the consistency of the modified gravity.

We may moreover note that the relation
\begin{equation}
\frac{p}{p_0}=\left( \frac{a}{a_0}\right)^{-3\gamma}, \label{27}
\end{equation}
known from ordinary relativity, is satisfied also here, for
arbitrary values of $\alpha$.

\section{Introduction of Viscosity}

Assume now that the fluid is viscous. Because of the assumed
spherical symmetry the shear viscosity plays no role, and only the
bulk viscosity $\zeta$ has to be considered. On thermodynamical
grounds, in conventional physics $\zeta$ has to be positive; this
being a consequence of the positive entropy change in irreversible
processes. We shall assume $\zeta >0$ also here. In general, the
presence of $\zeta$ does not have any influence upon the
(00)-component of the equations of motion. The only change in the
formalism because of viscosity is that the thermodynamical
pressure $p$ becomes replaced with the effective pressure $\tilde
p$, defined as
 \begin{equation} {\tilde p}=p-\zeta \theta \label{28}
\end{equation}
(cf., for instance, Refs.~\cite{brevik05,brevik94}). This follows
from the expression for the stress tensor for a viscous fluid. It
can be easily shown  that the $(rr)$-component of Eq.~(\ref{12})
(the pressure component of the equations of motion) becomes
satisfied for all $t$ if $\zeta$ is taken to be time dependent,
$\zeta \propto \theta^{2\alpha -1}$, or
\begin{equation}
\zeta =\frac{\zeta_0}{(1+\beta t)^{2\alpha-1}}. \label{29}
\end{equation}
Namely, the product $\zeta \theta$ then varies with $t$ in the
same manner as $p$,
\begin{equation}
\zeta \theta =\frac{\zeta_0 \theta_0}{(1+\beta t)^{2\alpha}};
\label{30}
\end{equation}
cf. Eq.~(\ref{24}). We do not need here to involve the pressure
equation explicitly, but can confine ourselves to the energy
equation (\ref{21}), making use of the equation of state
(\ref{2}). In the presence of viscosity the right hand side of
Eq.~(\ref{21}) thereby becomes changed to
\begin{equation}
\kappa^2\rho=\frac{\kappa^2}{\gamma-1}p \rightarrow
\frac{\kappa^2}{\gamma-1}(p-\zeta \theta). \label{31}
\end{equation}
Thus the generalized equation from which $\theta_0$ can be
determined in the viscous case, becomes
\begin{eqnarray}
\frac{1}{2}f_0 \,\left(\frac{\gamma}{\alpha}\right)^\alpha
\left(\frac{4\alpha}{3\gamma}-1\right)^{\alpha-1}
\Bigg\{(2-\alpha)\frac{2\alpha}{3\gamma}-(\alpha-1)(2\alpha-1)
\Bigg\}\theta_0^{2\alpha} \nonumber \\
=\kappa^2\left(\rho_0-\frac{\zeta_0 \,\theta_0}{\gamma-1}\right).
\label{32}
\end{eqnarray}
Let us here consider two special cases. As usual, the most
important case is that of Einstein's gravity, $\alpha=1,\,f_0=1$,
whereby Eq.~(\ref{32}) reduces to a quadratic equation in
$\theta_0$. We get
\begin{equation}
\theta_0=\frac{-3\kappa^2\zeta_0+\kappa^2\sqrt{9\,\zeta_0^2+12(\gamma-1)^2\rho_0/\kappa^2}}
{2(\gamma-1)}. \label{33}
\end{equation}
The case $\alpha=1$ turns out to have a bearing on the transition
of the fluid through the barrier $w=1$ into the phantom region. In
the notation of Ref.~\cite{brevik05} we can write in this case
\begin{equation}
\zeta=\tau \theta, \label{34}
\end{equation}
$\tau$ being a positive constant. As found in
Ref.~\cite{brevik05}, if the magnitude of the bulk viscosity is so
large that the condition
\begin{equation}
3\kappa^2 \tau > \gamma \label{35}
\end{equation}
is satisfied, then the fluid will be driven into the phantom
region ($w<-1$) and the Big Rip singularity, even it started out
in the quintessence region ($w>-1$). This result follows directly
from Friedmann's equations for a viscous fluid, and shows up
mathematically as a singularity in the density $\rho$ occurring at
a finite time $t$.

As our second example we will consider the case $\alpha=1/2$,
meaning that the gravitational action (\ref{1}) involves the
square root of the scalar curvature. Mathematically,
Eq.~(\ref{32}) then becomes quite simple and yields the solution
\begin{equation}
\theta_0=\frac{\kappa^2\,\rho_0}
{f_0\sqrt{\frac{3}{16(1-3\gamma/2)}}+\frac{\kappa^2\,\zeta_0}{\gamma-1}}.
\label{36}
\end{equation}
The initial scalar expansion becomes thus proportional to the
initial density. The same proportionality holds true for the time
dependent quantities $\theta$ and $\rho$; cf. Eqs.~(\ref{8}) and
(\ref{22}).

\section{Summary}

We started out with a spatially flat FRW space, adopting a
modified gravity action in the form of Eq.~(\ref{1}); this form
being recently studied in Ref.~\cite{abdalla05}. The corresponding
equations of motion are given in Eq.~(\ref{12}).
 We made the basic ansatz
 (\ref{6}) for the time variation of the scale factor $a$. This
 ansatz turned out to function well throughout, thus showing in
 general the flexibility and usefulness of the modified gravity
 formalism. We considered also the energy-momentum conservation,
 with similar satisfactory results.

 As far as viscosity is concerned, the formalism admits the
 presence of a bulk viscosity $\zeta$ without any problems if
 $\zeta$ is taken to be proportional to the scalar expansion
 $\theta$ raised to the power $(2\alpha-1)$; cf. Eq.~(\ref{29}).
  Then the product $\zeta\,\theta$
 varies with time in the same manner as the pressure $p$, and the
 equations of motion are satisfied. This behavior could hardly
 have been seen beforehand, without a detailed consideration of the formalism. In
 the special case of Einstein's gravity ($\alpha=1$) this result
 fits nicely into the viscous cosmology description recently given
 in Ref.~\cite{brevik05}: there exists in principle a
 viscosity-driven transition of the fluid from the
 quintessence region into the phantom region, implying a future Big Rip
singularity.

Finally we mention that our inclusion of bulk viscosity may be
used in more general modified gravities, for instance, those
introduced by Allemandi {\it et al.} \cite{allemandi05}.

\section*{Acknowledgments}

We thank Sergei Odintsov for many valuable comments on this
manuscript.


\end{document}